\documentclass{aa}  
\usepackage{graphicx}
\usepackage{txfonts,ulem}
\usepackage{mathrsfs}
\usepackage{color}
\usepackage{lscape}
\usepackage{xspace}

\newcommand{\Oi}{\Omega_{\mathrm{I}}}
\newcommand{\Oe}{\Omega_{\mathrm{E}}}
\newcommand{\Os}{\Omega_{\mathrm{S}}}
\newcommand{\kepler}{\textit{Kepler}\xspace}

\newcommand{\vitto}{\object{KIC004914923}\xspace}
\newcommand{\kitty}{\object{KIC005184732}\xspace}
\newcommand{\nunny}{\object{KIC006116048}\xspace}
\newcommand{\fred}{\object{KIC006933899}\xspace}
\newcommand{\rudy}{\object{KIC010963065}\xspace}
\newcommand{\pmodes}{\textit{p}-modes\xspace}
\newcommand{\pmode}{\textit{p}-mode\xspace}

\newcommand{\muHz}{\,\mu\mathrm{Hz}}

\begin{document} 

\title{Limits on radial differential rotation in Sun-like stars from parametric fits to oscillation power spectra}

\author{M.~B.~Nielsen\inst{1,2}
        \and
        H.~Schunker \inst{2}
        \and
        L.~Gizon \inst{2,3,1}
        \and
        J.~Schou \inst{2}
        \and
        W.~H.~Ball \inst{3,2,4}
        }

\institute{Center for Space Science, NYUAD Institute, New York University Abu Dhabi, PO Box 129188, Abu Dhabi, UAE\\
           \email{mbn4@nyu.edu}
           \and Max-Planck-Institut f{\"u}r Sonnensystemforschung, Justus-von-Liebig-Weg 3, 37077 G{\"o}ttingen, Germany
           \and Institut f{\"u}r Astrophysik, Georg-August-Universit{\"a}t G{\"o}ttingen, Friedrich-Hund-Platz 1, 37077 G{\"o}ttingen, Germany
           \and School of Physics and Astronomy, University of Birmingham, Edgbaston, Birmingham, B15 2TT, UK
           }

\date{Received XXXXXX XX, 2017; accepted XXXXXX XX, 2017}

\abstract
{Rotational shear in Sun-like stars is thought to be an important ingredient in models of stellar dynamos. Thanks to helioseismology, rotation in the Sun is  characterized well, but the interior rotation profiles of other Sun-like stars are not so well constrained. Until recently, measurements of rotation in Sun-like stars have focused on the mean rotation, but little progress has been made on measuring or even placing limits on differential rotation.}
{Using asteroseismic measurements of rotation we aim to constrain the radial shear in five Sun-like stars observed by the NASA \kepler mission:  \vitto, \kitty, \nunny, \fred, and \rudy.}
{We used stellar structure models for these five stars from previous works. These models provide the mass density, mode eigenfunctions, and the convection zone depth, which we used to compute the sensitivity kernels for the rotational frequency splitting of the modes. We used these kernels as weights in a parametric model of the stellar rotation profile of each star, where we allowed different rotation rates for the radiative interior and the convective envelope. This parametric model was incorporated into a fit to the oscillation power spectrum of each of the five \kepler stars. This fit included a prior on the rotation of the envelope, estimated from the rotation of surface magnetic activity measured from the photometric variability.}
{The asteroseismic measurements without the application of priors are unable to place meaningful limits on the radial shear. Using a prior on the envelope rotation enables us to constrain the interior rotation rate and thus the radial shear. In the five cases that we  studied, the interior rotation rate does not differ from the envelope by more than approximately $\pm 30 \%$. Uncertainties in the rotational splittings are too large to unambiguously determine the sign of the radial shear.}
{}

\titlerunning{Limits on radial differential rotation in Sun-like stars}
\authorrunning{M.~B.~Nielsen et al.}
\maketitle

\newpage
\section{Introduction}
Helioseismology has shown that the Sun exhibits a complex differential rotation profile. The radiative interior rotates as a solid body, while the convective envelope predominantly shows a decreasing rotation rate with increasing latitude \citep{Schou1998}. Rotation in the convection zone is largely constant with radius, except at the surface where a thin shear layer exists, and at the interface with the radiative interior (known as the tachocline). Differential rotation like that seen in the Sun may sustain the solar magnetic dynamo. The relative importance of each of these regions for generating the magnetic activity is, however, not clearly understood  \citep[see, e.g.,][for a review]{Charbonneau2010}. Models that seek to explain the solar dynamo must necessarily be limited by the scale of the differential rotation in the Sun, but must also be robust against estimates of differential rotation and activity measurements of other stars.

Measurements of latitudinal differential rotation on other stars has been possible for some time  using high-resolution spectroscopy \citep{Reiners2003}, Doppler imaging \citep{Vogt1983,Collier2002,Hackman2012}, spectropolarimetric data \citep{Semel1989,Donati1997,Jeffers2014};  variations of the rotation periods measured by photometric variability have also been used \citep{Reinhold2013,Lehtinen2016}.

Radial differential rotation, on the other hand, is more difficult to measure since the only means of studying stellar interiors is through asteroseismology. Stars with a sufficiently strong driving mechanism (e.g.,  convective motion in the case of solar-mass stars or radial opacity variations in more massive stars) are able to excite oscillations that propagate through the stellar interior \citep[see, e.g.,][]{Aerts2010}. These oscillations are perturbed by velocity fields such as rotation in the interior, and so the frequencies of the brightness variations seen on the surface are similarly perturbed. However, high signal-to-noise, long duration observations are necessary to measure these minute perturbations, especially in Sun-like stars where the sensitivity of the modes to rotation in the deep interior is lower than in the envelope. 

Radial differential rotation has been measured in a selection of red giant stars \citep{Beck2012,Deheuvels2012,Deheuvels2015}, and in cool  subgiant stars \citep{Deheuvels2014}. Such stars exhibit gravity-dominated mixed modes, which are sensitive to rotation in the deep interior. In addition, \citet{Benomar2015} studied a sample of predominantly fast rotating F-type stars, where they combined asteroseismic measurements of the mean stellar rotation rate with spectroscopic $v\sin{i}$. Differences between the two were attributed to differential rotation between the radiative interior and convective envelope. Here we focus on Sun-like main-sequence stars, a population which has so far remained unexplored  with respect to radial differential rotation.

In stars like the Sun only the pressure dominated \pmodes are visible at the surface, and these have very little sensitivity to the rotation in the radiative interior. This, in combination with their typically slow rotation rate \citep{McQuillan2014}, makes robust estimates of radial differential rotation difficult to obtain \citep[see, e.g.,][]{Lund2014}. Approaches like those used for red giants and subgiants are therefore unable to place meaningful constraints on the radial differential rotation \citep{Schunker2016a}. Furthermore, because of their slow rotation the errors on spectroscopic $v \sin{i}$ measurements of Sun-like stars become prohibitively large, making the approach used by \citet{Benomar2015} difficult. However, surface rotation measurements obtained from photometric variability can be used in a similar fashion to constrain the surface rotation. 

In \citet{Nielsen2015} we compared the average internal rotation rate with that derived from surface variability, for a small sample of slowly rotating Sun-like stars (listed in Table~\ref{tab:results}).  This showed that the differential rotation between the surface and the interior is likely very small. Here, we use the surface rotation rate as a prior on the rotation of the convective envelope, in a direct asteroseismic fit to the power spectra of the same stars. This fit incorporates a two-parameter model for the radial rotation profile, consisting of separate rotation rates of the radiative interior and convective envelope. The surface rotation rate, which is derived from photometric variability from active regions, is then used to constrain the rotation of the convective envelope. The approach adopted in the present work improves upon previous measurements in the sense that it allows for variation in the sensitivity of the modes to rotation at different depths inside the star, combined with surface rotation measurements, thus providing tighter constraints on the radial shear at the tachocline.

\section{Rotation from surface activity}\label{sec:surfrot}
We use measurements of rotation from surface variability originally performed in \citet{Nielsen2015}, which are listed in Table~\ref{tab:results}. The data characteristics and method of measuring rotation will therefore only be briefly covered here.

For both the measurements of the surface rotation and the asteroseismic analysis we used white-light photometric observations from NASA's \kepler satellite \citep{Borucki2010}. These observations span approximately $1400$ days, and are available with two different integration times: long cadence (LC) at $\sim 29.45$ minutes, and short cadence (SC) at $\sim 59$ seconds. These observations are made available through the Mikulski Archive for Space Telescopes\footnote{https://archive.stsci.edu/}. 

To measure surface rotation we use the LC time series, which were pre-processed for systematic noise using the automated PDC\_MAP and msMAP pipelines \citep{Stumpe2012,Smith2012}. However, we also performed manual time series extraction and reduction from the raw pixel level data using the software package PyKe\footnote{http://keplerscience.arc.nasa.gov/PyKE.shtml}. This was done to ensure that any variability observed in the time series was not caused by the automated reduction pipelines \citep[see][for details]{Nielsen2015}. 

For each quarter of \kepler observations, we infer the period of rotation by identifying the peak with highest amplitude in the Lomb-Scargle periodogram. The average rotation period over the full duration of the \kepler observations, which  here we call $\Pi_\mathrm{S}$, is then obtained by taking the median of the rotation periods over all available quarters. The associated standard deviation, $\sigma_{\mathrm{S}}$, is obtained by $1.48 \cdot MAD$, where MAD is the median absolute deviation of the rotation periods over the available quarters \citep{Nielsen2015}. In Table~\ref{tab:results} we show the values of $\Os=2\pi/\Pi_{\mathrm{S}}$, with asymmetric errors corresponding to the $68\%$ confidence interval.

\section{Rotation from asteroseismology}
In this section we describe the data and methods used to extract rotation information from the stellar oscillations. 

The oscillations of Sun-like stars have periods on the order of 5-10 minutes, and we therefore use the SC time series for this part of the analysis. We use the power density spectra of these observations that have been made available through the Kepler Asteroseismic Science Operations Center\footnote{http://kasoc.phys.au.dk/} (KASOC). Prior to computing the power density spectra, the time series were processed using the KASOC filter \citep{Handberg2014}, which optimizes them for asteroseismic analysis. 

This filtering includes removing any possible planetary transits and variability caused by surface features on the stars, as well as discontinuities between quarters. Such features and variability may elevate the noise level in the frequency range containing stellar oscillation modes, thereby potentially obscuring the minute effects of rotation.

\subsection{Maximum likelihood estimation}
The maximum likelihood approach is a method to fit a parametric model $M(\boldsymbol{\Theta},\nu_j)$ of the expectation value of the power spectrum to a realization $\{P_j\}$ of the power spectrum. The components of the vector $\boldsymbol{\Theta}$ are the parameters to be determined in the fit. The observed power $P_j$ at frequency $\nu_j$ is distributed according to a $\chi^2$ distribution with 2 degrees of freedom \citep[e.g.,][]{Woodard1984,Duvall1986,Gizon2003}. The probability of observing a given value $P_j$ in the power spectrum is thus an exponential distribution:
\begin{equation}
f_j\left(P_j, \boldsymbol{\Theta} \right) = \frac{1}{M(\boldsymbol{\Theta},\nu_j)}\mathrm{exp}\left({-\frac{P_j}{M(\boldsymbol{\Theta},\nu_j)}}\right).
\end{equation}

For high duty-cycle observations like those from \kepler ($\gtrsim 90\%$) the power in two different frequency bins can be assumed to be independent \citep[e.g.,][]{Stahn2010}. Hence, the joint probability of the observations is given by the product of the individual probabilities $f_j$, where $j$ spans the frequency domain of interest (in this paper the whole power spectrum). The best-fit parameters $\boldsymbol{\hat{\Theta}}$ are the parameters that maximize the log-likelihood, which is the logarithm of the joint probability function evaluated at the observed $\{P_j\}$,
\begin{equation}
 L\left(\boldsymbol{\Theta} \right) = \ln \prod\limits_{j=1}^J f_j\left(P_j,\boldsymbol{\Theta}\right) = - \sum\limits_{j=1}^J {\left(\ln M\left( {\boldsymbol{\Theta} ,\nu _j } \right) + \frac{{P_j }}{{M\left( {\boldsymbol{\Theta} ,\nu _j } \right)}}\right)},
\end{equation}
where $J\sim 10^6$ is the number of frequencies. Here we use a Markov chain Monte Carlo sampler\footnote{We use the Python package EMCEE, available at http://dan.iel.fm/emcee/current/} \citep{Foreman-Mackey2013} to search the parameter space and to return the best-fit estimate $\boldsymbol{\hat{\Theta}}$. We note that the logarithm of the joint probability is used for better numerical stability in the optimization.

\begin{figure*}
\centering
\includegraphics[width = 2\columnwidth]{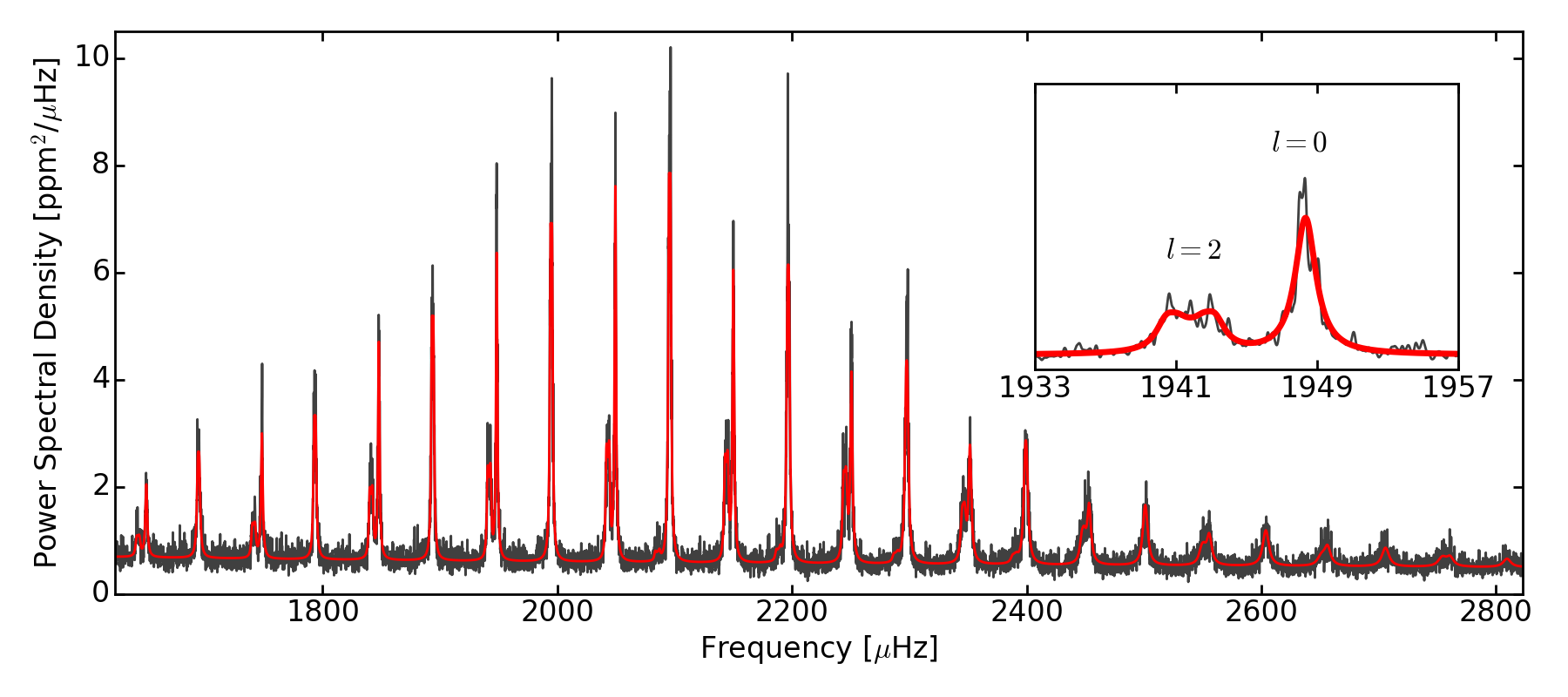}
\caption{Smoothed power spectrum of \nunny shown in black, and the best-fit model shown in red. The inset shows a zoom around $1945\muHz$ at an $l = 2, 0$ pair. The $l = 2$ consists of a multiplet of $2l + 1$ azimuthal orders with frequencies that are separated by the effects of rotation. In this case the rotation rate is not high enough to separate the modes by more than the intrinsic broadening caused by the damping of the mode. The $l = 2$ multiplet is therefore only broadened rather than clearly split into its individual azimuthal components.}
\label{fig:spectrum}
\end{figure*}

\subsection{Power spectrum model}
The pulsations in a Sun-like star are stochastically excited and damped by the convective motion in the outer layers of the star. The oscillations are decomposed into spherical harmonic functions with angular degree $l$ and azimuthal order $m$, as well as a radial order $n$ which describes the number of nodal shells in the radial direction. 

Figure~\ref{fig:spectrum} shows the spectrum of \nunny and a best-fit model. The inset shows an $l=2,0$ pair, where the $l=2$ peak is a multiplet consisting of $2l+1$ rotationally-split azimuthal components. The $l=1$ (not shown in the inset) and $l=2$ multiplets contain the rotation information.

The expectation value of the power of a single oscillation mode can be approximated by a Lorentzian profile \citep{Anderson1990}. The oscillation spectrum of a Sun-like star contains a number of these modes, and we therefore model the spectrum as a sum of Lorentzian profiles
\begin{equation}
M\left( {\boldsymbol{\Theta} ,\nu } \right) = \sum\limits_n^{} {\sum\limits_{l = 0}^3 {\sum\limits_{m = - l}^l {\frac{{S_{nl} \, \mathscr{E}_{lm} \left( i \right) 
}}{{1 + \left( {2/\Gamma _{nlm} } \right)^2 \left( {\nu - \nu _{nlm} } \right)^2 }} + B\left( \nu \right)} } }. 
\end{equation}
Each mode has a peak height of $S_{nl} \mathscr{E}_{lm} \left( i \right)$, a full width at half maximum $\Gamma_{nlm}$, and a central frequency $\nu_{nlm}$. Because of the long duration of the \kepler observations, the mode width $\Gamma_{nlm}$ is dominated by the damping rate of the oscillations by convective motions. The damping rate, and therefore the mode width, increases smoothly with frequency in a Sun-like star. We therefore parameterize the mode width by a $5$th order polynomial \citep{Stahn2010}.

The parameter $S_{nlm}$ is a function of the oscillation amplitude and the instrument-dependent visibility of a given mode. These are not parameters of interest for the analysis of rotation performed here and they are therefore combined in $S_{nl}$, which is left as a free parameter for each multiplet in the model fit.

An additional modulation $\mathscr{E}_{lm} \left( i \right)$ of the peak height comes from the inclination angle, $i$, of the stellar rotation axis relative to the line of sight to the observer. We use the same expression for $\mathscr{E}_{lm} \left( i \right)$ as given by \citet{Gizon2003}. This accounts for the relative differences in peak height of the azimuthal components with the same radial order and angular degree, and allows us to determine the stellar inclination angle from the power spectrum fits.

The mode frequencies $\nu_{nlm}$ are perturbed by rotation
\begin{equation}
\nu_{nlm} = \nu_{nl} + m \int\limits_0^{\pi}{\int\limits_0^{R}{\frac{\Omega(r,\theta)}{2\pi}K_{nlm}(r,\theta) r dr d\theta}},
\label{eq:mode_frequency}
\end{equation}
where $\nu_{nl}$ is the average mode frequency of a multiplet $nl$, $\Omega(r,\theta)$ is the internal angular velocity of the star, and $K_{nlm}(r,\theta)$ is the rotation sensitivity kernel for the mode $nlm$. The kernels give the sensitivity of the modes to the local angular velocity in the star \citep[see][Eq. (8.35)]{JCDoscnotes}, and depend on the structure of the star. These kernels are computed using stellar structure models originally from \citet{Nielsen2015}. 

In addition to the oscillation modes the power spectrum also contains a level of background noise. This is modeled by  
\begin{equation}
B\left( \nu \right) = \sum\limits_{q = 1}^Q {\frac{{A_q \tau _q }}{{1 + \left( {2\pi \nu \tau _q } \right)^{\alpha _q } }}} + W,
\end{equation}
which consists of a frequency-independent shot noise level $W$ and a level of frequency-dependent red noise. The frequency-dependent noise is caused primarily by the stochastic variability of the granulation on the stellar surface, and by the very low-frequency noise from surface activity and instrumental effects \citep{Harvey1988}. The power excess from these noise sources is modeled using $Q$ Harvey-like terms \citep[see, e.g.,][]{Aigrain2004,Kallinger2014}, which have a characteristic amplitude $A_q$, time-scale $\tau_q$, and decay with increasing frequency determined by the exponent $\alpha_q$. For most of the stars $Q=2$ Harvey-like terms are sufficient, but for \vitto an additional term ($Q=3$) located at $\sim 1200\mu$Hz is required to adequately account for the background variation with frequency. Similar noise terms are marginally visible in the residuals of the other stars, but not to the same extent as for \vitto. We do not expect the omission of this term to influence the rotation measurements for these stars. \citet{Karoff2013} suggest that the source of this noise may be faculae on the stellar surface, with variability on shorter timescales than the more clearly visible granulation pattern at $\sim100-400 \mu$Hz. 

\subsection{Representing the radial shear by a step function}
It is currently not possible to obtain resolved estimates of $\Omega(r,\theta)$ for stars other than the Sun by inversion of the mode frequencies \citep[][]{Lund2014,Schunker2016a}. It is possible, however,  to test how well the data can constrain an assumed model rotation profile.

The solar rotation profile has a shear layer at the base of the convection zone between the uniformly-rotating interior and the convective envelope. While the rotational shear is both radial and latitudinal in the Sun, in this paper we choose to constrain only the radial shear, effectively averaging over the latitudinal component of the differential rotation. We parameterize the angular velocity profile of the five Sun-like stars under study by a simple radial step function in the model fit
\begin{equation}
\Omega \left( r \right) = \left\{ {\begin{array}{*{20}c}
   {\Oi\quad \mathrm{for}} & {0 \le r < r_{cz} } \\
   {\Oe\quad \mathrm{for}} & {r_{cz} \le r \le R} , \\
\end{array}} \right. 
\label{eq:rotation_profile}
\end{equation}
where $R$ and $r_{cz}$ are the radius of the star and the base of the convection zone, respectively. Both $R$ and $r_{cz}$ are determined from stellar structure models in \citet{Nielsen2015}. The two parameters $\Oi$ and $\Oe$ are the rotation rates of the radiative interior and convective envelope, and a significant difference between the two would indicate the presence of shear at the base of the convection zone.

With this choice of rotation profile, Eq.~\ref{eq:mode_frequency} can be rewritten as
\begin{equation}
\nu_{nlm} = \nu_{nl} + m\frac{\Oi}{2\pi} \int\limits_0^{r_{cz}}{K_{nl}(r)dr}+m\frac{\Oe}{2\pi} \int\limits_{r_{cz}}^{R}{K_{nl}(r)dr},
\label{eq:nunlm}
\end{equation}
where the kernels $K_{nl}$ are functions of radius only. The kernels depend on the unperturbed eigenfunctions and mass density from the stellar structure model, which are computed prior to the fit to the oscillation power spectrum.
In Eq.~\ref{eq:nunlm} the only free variables are therefore $\nu_{ln}$, $\Oi$, and $\Oe$.

The error on the stellar radius $R$ is approximately $1\%$, and is estimated to be at a similar level for the convection zone radius because of the  high-quality data. Variations of this scale on the radii are not expected to contribute significantly to the estimate of the radial shear \citep{Schunker2016}, and we therefore keep both radii fixed in the model fits. These values are listed in Table~\ref{tab:results}.

\begin{figure}
\centering
\includegraphics[width = \columnwidth]{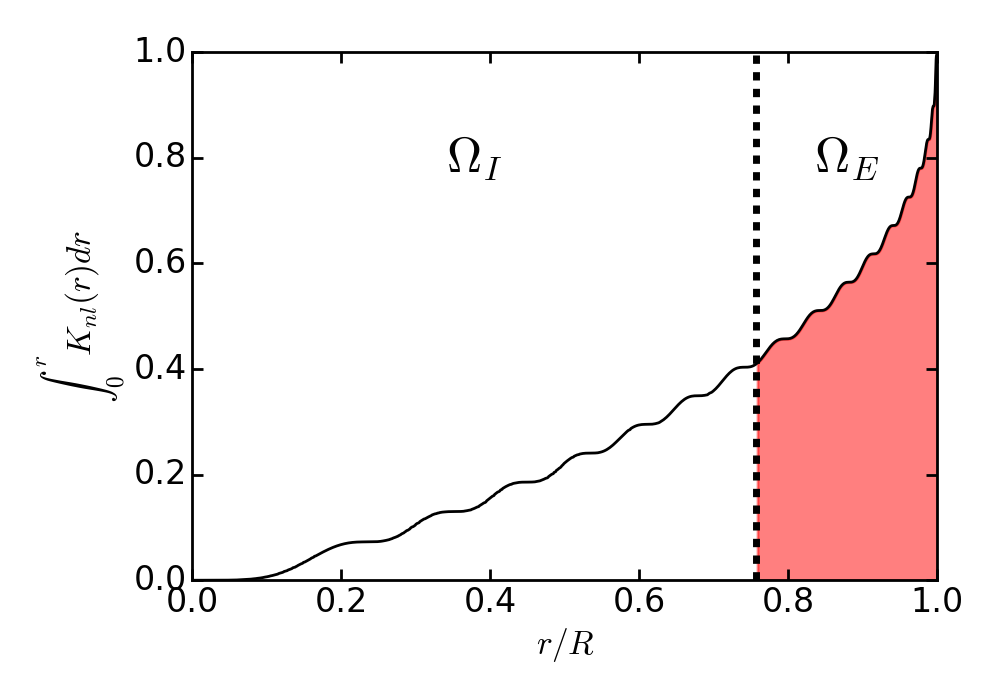}
\caption{Cumulative integral of the radial component of the rotation sensitivity kernel of the $n=17,l=1$ mode in \nunny. The shaded area denotes the part of the integral which is sensitive to the envelope rotation $\Oe$, while the unshaded part is sensitive to the interior rotation $\Oi$. The dashed line denotes the radius of the base of the convection zone $r_{cz}$ which separates the two parts of the star.}
\label{fig:radial_kernel}
\end{figure}

Figure~\ref{fig:radial_kernel} shows the cumulative integral of the radial sensitivity kernel $K_{nl}(r)$ for the $n=17$, $l=1$ mode in \nunny. The dashed line marks the base of the convection zone at a radius of $r_{cz} =0.757R$, separating the convective envelope (shaded red) and the radiative interior. We note that the integral of the kernel is never exactly unity, due to the effect of the Coriolis force \citep{Ledoux1951}. The cumulative integral up to the base of the convection zone reaches a value of $\sim 0.41$, showing that the oscillation modes are still moderately sensitive to rotation below the convection zone.

Figure~\ref{fig:spectrum} shows the model fit to the power spectrum of \nunny. The perturbations to the mode frequencies are only visible on scales of a few $\mu$Hz, as shown by the splitting of the $l=2$ mode in the inset. The posterior distributions of $\Oe$ and $\Oi$ are shown in Fig.~\ref{fig:correlation}. The large widths of the posterior density distributions of $\Oe$ and $\Oi$ indicate that the information from the oscillation modes alone is not able to constrain the two parameters, allowing a wide range of likely solutions for either. However, it is clear that the two parameters are very strongly anti-correlated. Therefore, if a prior is applied to one parameter the other may also be constrained. 

The results of the initial fits without priors are given in the middle section of Table~\ref{tab:results};  all the stars show similar rotation rates corresponding to $\Oe \approx \Oi$ within the $68\%$ confidence interval. An exception to this is \vitto where the best-fit values suggest $\Oi-\Oe = -8.84_{-5.84}^{+5.59}\,\mu$Hz. While the $95\%$ confidence interval (corresponding to $2\sigma$) extends to encompass $\Oe=\Oi$ as for the other stars, it would appear that the most likely configuration for \vitto is the interior and envelope spinning rapidly in opposite directions. However, this deviation is likely caused by the strong anti-correlation between the inclination angle and $\Oe$ (and subsequent correlation with $\Oi$). To test this, we fixed $i$ to its $84$th percentile value (see Table~\ref{tab:results}, middle section) and re-sampled the posterior distributions of $\Oe$ and $\Oi$. We found that these solutions correspond to a configuration with $\Oe \approx \Oi$.

The fit to \vitto is different from the fit to the other stars only in that it was necessary to fit a third  Harvey-like background term, but otherwise the spectrum and model terms are identical to the other stars. We searched for potential causes of the deviation in the inclination angle by running a fit for \vitto with only $Q=2$ background terms as for the other stars, using only the high signal-to-noise modes near the center of the \pmode envelope, and adding additional marginally visible modes at higher and lower frequencies. However, none of these approaches produced a significantly different value of the inclination angle. In addition, we note that the best-fit rotation rates do not change appreciably whether the $\ell=3$ modes are included in the model or not.

\section{Limits on radial differential rotation}

\subsection{Adding a prior from surface rotation}
\begin{figure*}
\centering
\includegraphics[width = 1.5\columnwidth]{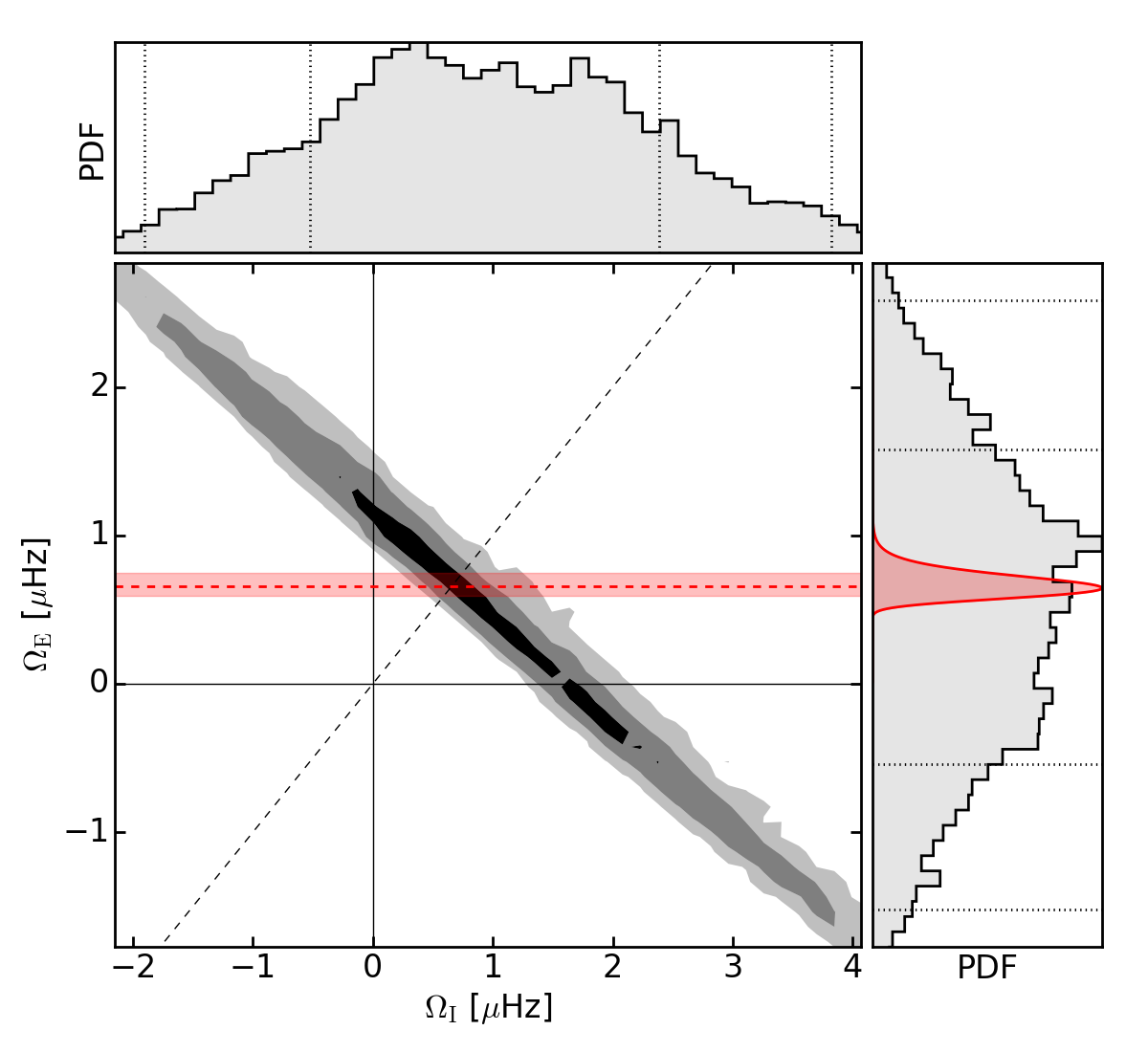}
\caption{Correlation between the posterior density of $\Oe$ (right) and $\Oi$ (top) for \nunny before the application of a prior (in grayscale). From the posteriors themselves it is clear that the two parameters are poorly constrained, with a standard deviation of several $\mu$Hz in both cases. The two parameters are, however, strongly anti-correlated. In the center frame the dashed black line indicates $\Oe = \Oi$. Dashed red denotes the median of the $\Os$ probability density function, and the shaded bar corresponds to the $68\%$ confidence interval. The full distribution of $\Os$ is shown in the right panel, which is used as a prior to constrain $\Oe$.}
\label{fig:correlation}
\end{figure*}

In the following we assume that the rotation period, $\Os$, from surface variability provides a constraint on the rotation period of the stellar envelope. In practice, we assume that the prior on $2\pi/\Oe$ is Gaussian with a mean $2\pi/\Os$ and standard deviation $\sigma_{\mathrm{S}}$, as defined in section~\ref{sec:surfrot}. The log-likelihood function that must be maximized then becomes
\begin{align}
  L\left( {\boldsymbol{\Theta}} \right) =& \begin{aligned}[t]
      &- \sum\limits_{j = 1}^J {\left( {\ln M\left( {\Theta ,\nu _j } \right) + \frac{{P_j }}{{M\left( {\boldsymbol{\Theta},\nu _j } \right)}}} \right)}\\
      &- \frac{1}{{2\sigma _{\mathrm{S}}^2 }}\left( {\frac{{2\pi }}{{\Omega _E }} - \frac{{2\pi }}{{\Omega _S }} } \right)^2.
       \end{aligned}
\end{align}

Figure~\ref{fig:correlation} shows an example of the prior on $\Oe$ in red for \nunny. Given the asteroseismic measurements, the prior on $\Oe$ provides a constraint on $\Oi$.

\subsection{Constraints on the radial shear}
Table~\ref{tab:results} lists the best-fit values of $\Oe$ and $\Oi$ with and without the application of a prior. Comparing $\Oe$ before and after the use of the prior, we find that $\Oe$ matches $\Os$ to a high degree. This means that the prior dominates the fit for $\Oe$, and the spectra do not appear to contain any significant information to constrain $\Oe$ when $\Oi$ is also a free parameter. Because of the strong correlation between the two parameters, this constraint on $\Oe$ also reduces the range of likely values of $\Oi$. The median $\Oi$ values consistently fall close to those of $\Oe$ for all the stars, with errors between $\sim0.08-0.36$~$\mu$Hz. This uncertainty stems from the relatively low sensitivity to rotation of the modes in the interior, and is the main contributor to the error on the radial shear. 

The parameter of interest is the difference between the two rotation rates, and are shown in the lower section of Table \ref{tab:results}. The median values of $\Oe$ and $\Oi$ with the prior included suggest that the most likely configuration for all the stars is close to $\Oe=\Oi$, corresponding to solid-body rotation. The error on $\Oi-\Oe$ is now on the order of $\lesssim 0.4 \mu$Hz (compared to $>1\mu$Hz without the prior), indicating that shear values are typically very small and likely do not exceed a few hundred nHz.

For comparison between the stars we also list the relative differences between the envelope and interior rotation rates in Table~\ref{tab:results}. Figure~\ref{fig:rel_diff} shows the probability density distributions of these relative differences after the application of the priors in the fit. The distributions predominantly have median values close to zero and are roughly symmetric (except \kitty, see below). The median of the $68\%$ confidence interval for all the stars range from $-29\%$ to $+34\%$. Because of the near symmetry it is not possible to infer which rotates faster in Sun-like stars, the radiative interior or the convective envelope.

\subsection{Special case of \kitty}
\label{sec:kitty}
\begin{figure*}
\centering
\includegraphics[width = 2\columnwidth]{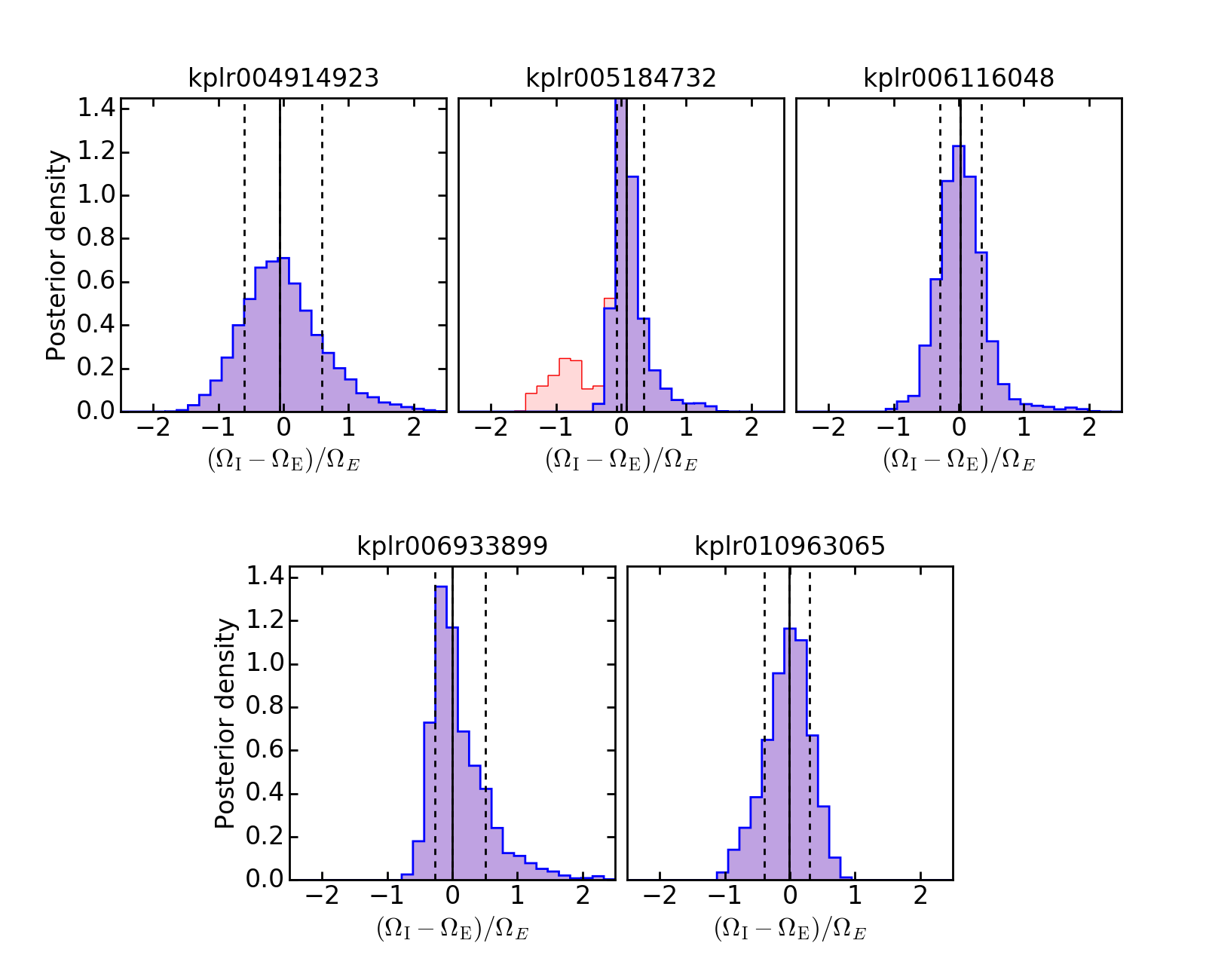}
\caption{\label{tab:results}
Relative differences between the envelope and interior rotation rates for each of the stars in the analyzed sample. For \kitty the red shaded part of the posterior denotes the values where $\Oe$ is exceptionally high and so a likely  alias of the real rotation rate (see section~\ref{sec:kitty}). Solid and dashed black vertical lines denote the median (16th and 84th percentiles of the distributions);  for \kitty they apply only to the dark shaded part of the posterior distribution.}
\label{fig:rel_diff}
\end{figure*}

The posterior of $\Oe$ for most of the stars is similar to the corresponding prior distribution. However, for \kitty this is not the case. The posterior density shows a second maximum at $\Oe/2\pi = 1.33_{-0.24}^{+0.32} \muHz$, separated by a low-likelihood region from the location of the prior at $\Oe/2\pi = 0.56 \pm 0.01 \muHz$. This local maximum gives rise to the bimodality indicated by the red shaded region in Fig.~\ref{fig:rel_diff} for \kitty. The reason for this is illustrated in Fig.~2 of \citet{Gizon2003}, which shows that for inclinations of $60^{\circ}$ to $70^{\circ}$ the peak height of the $m=\pm1$ components are comparable to the $m=\pm2$ components of the $l=2$ multiplet. This ambiguity in peak height gives rise to a secondary maximum at close to twice the rotation rate of the star since the azimuthal orders are separated approximately by integer multiples of the rotation rate. Since \kitty has an inclination of $\sim 71^{\circ}$ (see Table~\ref{tab:results}) this ambiguity is expected. The rotation rate is, however, also strongly constrained by the $l=1$ multiplet, which reduces the significance of this local maximum. Because this maximum is considered an alias of the true rotation rate, it is omitted when computing the results for \kitty shown in Table~\ref{tab:results}.

This same feature is also visible for the other stars, but to a much lesser extent. In these cases the separation between the global and local maximum is much less clear. Since they do not appear to contribute significantly to the estimation of $\Oe$ and $\Oi$, they are not removed. 

\begin{table*}
\caption{Fundamental stellar parameters (top) and best-fit values for $\Oe$ and $\Oi$, without (middle) and with (bottom) a prior on $\Oe$.}
\centering
\begin{tabular}{lccccc}
\hline\hline
                              & \object{KIC004914923}  & \object{KIC005184732}  & \object{KIC006116048}  & \object{KIC006933899}  & \object{KIC010963065} \\
\hline
$\mathrm{T_{eff}}$[K]         & $5807\pm89$            & $5744\pm63$            & $5968\pm58$            & $5764\pm46$            & $6158\pm57$ \\
$M/\mathrm{M}_{\odot}$        & $1.118\pm0.020$        & $1.205\pm0.025$        & $1.023\pm0.021$        & $1.096\pm0.026$        & $1.062\pm0.021$ \\
$R/\mathrm{R}_{\odot}$        & $1.378\pm0.009$        & $1.342\pm0.010$        & $1.225\pm0.008$        & $1.574\pm0.025$        & $1.220\pm0.009$ \\
$r_{cz}/R$                    & $0.724$                & $0.731$                & $0.757$                & $0.786$                & $0.714$ \\
$[\mathrm{Fe}/\mathrm{H}]$    & $0.091\pm0.083$        & $0.384\pm0.025$        & $-0.191\pm0.049$       & $0.126\pm0.037$        & $-0.189\pm0.038$ \\
Age [Gyr]                     & $6.23\pm0.36$          & $4.39\pm0.13$          & $5.70\pm0.21$          & $6.57\pm0.30$          & $4.18\pm0.19$ \\
$\Os$ [$\mu$Hz] & $0.63_{-0.09}^{+0.13}$ & $0.56_{-0.01}^{+0.01}$ & $0.66_{-0.07}^{+0.09}$ & $0.37_{-0.02}^{+0.02}$ & $0.94_{-0.02}^{+0.03}$ \\
\hline
Without prior       & & & & & \\
$\Oe$ [$\mu$Hz] & $4.11_{-2.19}^{+2.28}$  & $1.14_{-0.88}^{+0.92}$  & $0.79_{-0.75}^{+0.74}$     & $0.29_{-1.41}^{+1.63}$    & $1.28_{-1.79}^{+2.01}$  \\
$\Oi$ [$\mu$Hz] & $-4.73_{-3.55}^{+3.40}$ & $-0.07_{-1.30}^{+1.34}$  & $0.50_{-1.04}^{+1.04}$    & $0.49_{-2.08}^{+1.83}$    & $0.28_{-3.13}^{+2.99}$  \\
$(\Oi-\Oe)$ [$\mu$Hz]     & $-8.84_{-5.84}^{+5.59}$ & $-1.22_{-2.21}^{+2.23}$  & $-0.28_{-1.78}^{+1.77}$   & $0.21_{-3.71}^{+3.25}$    & $-1.00_{-5.11}^{+4.75}$ \\
$i$ [deg]       & $36.24_{-4.31}^{+5.31}$ & $67.51_{-10.56}^{+9.42}$ & $72.63_{-10.25}^{+10.83}$ & $61.68_{-14.74}^{+18.42}$ & $42.76_{-4.83}^{+8.14}$ \\
\hline
With prior       & & & & & \\
$\Oe$ [$\mu$Hz] & $0.65_{-0.10}^{+0.13}$ & $0.56_{-0.01}^{+0.01}$ & $0.65_{-0.06}^{+0.09}$ & $0.37_{-0.02}^{+0.02}$ & $0.94_{-0.02}^{+0.03}$ \\
$\Oi$ [$\mu$Hz] & $0.60_{-0.31}^{+0.31}$ & $0.61_{-0.08}^{+0.14}$ & $0.65_{-0.14}^{+0.15}$ & $0.37_{-0.09}^{+0.18}$ & $0.93_{-0.36}^{+0.29}$ \\
$(\Oi-\Oe)$ [$\mu$Hz] & $-0.04_{-0.40}^{+0.38}$ & $0.05_{-0.09}^{+0.14}$ & $0.01_{-0.22}^{+0.19}$ & $0.00_{-0.10}^{+0.18}$ & $-0.01_{-0.37}^{+0.29}$ \\
$(\Oi-\Oe)/\Oe$ & $-0.06_{-0.54}^{+0.64}$ & $0.09_{-0.16}^{+0.26}$ & $0.02_{-0.31}^{+0.33}$ & $0.00_{-0.26}^{+0.50}$ & $-0.01_{-0.39}^{+0.31}$ \\
$i$ [deg] & $41.27_{-5.57}^{+7.73}$ & $71.62_{-10.89}^{+9.59}$ & $75.02_{-9.31}^{+9.01}$ & $67.53_{-17.08}^{+15.59}$ & $43.32_{-4.70}^{+7.91}$ \\
\end{tabular}
\tablefoot{Top: The effective temperature $T_{\mathrm{eff}}$, mass $M$, radius $R$, metallicity $[\mathrm{Fe}/\mathrm{H}]$, age, and radius of the tachocline $r_{\mathrm{cz}}$ are derived from stellar structure models. These structure models are obtained from fits to the oscillation mode frequencies and spectroscopic measurements of the effective surface temperature from \citet{Bruntt2012}. The surface rotation $\Os$ is computed from multiple measurements of rotation over the \kepler mission lifetime. All values are adapted from \citet{Nielsen2015}. Middle: Best-fit values for $\Oe$, $\Oi$, and the angle of the stellar inclination angle $i$, without the application of a prior. Bottom: Best-fit values for $\Oe$, $\Oi$, and $i$ with the PDF of $\Os$ used as a prior on $\Oe$. For all rotation rates and the inclination angle $i$ the errors are given by the 16th and 84th percentile values of the posterior densities of each parameter. For the relative difference between $\Oe$ and $\Oi$, the average of the percentiles is between $-29\%$ and $+34\%$, indicating the range of likely configurations of the differential rotation.}
\end{table*}

\section{Conclusion}
We fit the oscillation spectra of five Sun-like stars in the \kepler field, using a model of the stellar rotation profile with rotation rates $\Oi$ for the radiative interior and $\Oe$ for the convective envelope, with a shear at the base of the convection zone. 

The \kepler mission provides high-cadence, long-duration observations with high photometric precision. However, using the oscillation spectra alone it is not possible to constrain the difference between the two parameters in the rotation models. The posterior distributions of $\Oe$ and $\Oi$ from the unconstrained fit span a wide range of rotation rates, up to several times that estimated from surface variability. The unconstrained fit to \vitto even produces a configuration of opposite sign and rapid rotation for the interior and the envelope. 

The interior and envelope rotation rates are strongly  anti-correlated, which stems from the fact that the weighted average of the rotation rates are very well constrained by the asteroseismic rotational splittings. By applying a tight prior to the envelope rotation $\Oe$, we can obtain a constraint on the interior rotation rate $\Oi$. The prior is given by the surface rotation rate inferred from the photometric variability caused by surface activity \citep{Nielsen2015}. Using all available measurements of rotation to help constrain the interior differential rotation offers a clear advantage. Such a prior assumes that the mean rotation rate of the envelope matches that of the surface to within the uncertainties, i.e., that latitudinal differential rotation is negligible.

Using a parametric fit with priors to the power spectra of the five stars studied here, we were able to constrain the radial differential rotation to an average range of $-29\%$ to $+34\%$ of the surface rotation (see Table \ref{tab:results}). This is consistent with a previous study of Sun-like and F-type dwarfs by \citet{Benomar2015}, who found the likely range extending from $-41\%$ to $+54 \%$ using a different parameterization of the oscillation power spectrum. While the average range that we find is also biased toward higher values of $\Oi$ relative to $\Oe$, \rudy shows the opposite behavior. It is therefore not possible to unambiguously determine the sign of the differential rotation based on our sample.

These measurements constrain the radial differential rotation in our sample of Sun-like stars to less than a few hundred nHz. For comparison, the radial differential rotation in the Sun is on the order of $10$~nHz \citep{Schou1998}. Measuring solar-like differential rotation rates is not yet possible for other stars, even with the high-quality data from the \kepler mission. Asteroseismic rotation measurements of distant stars are limited by the relatively low number of observable modes, and by the large width of the modes relative to the small rotational frequency splitting typical of most Sun-like stars. However, these measurements still serve to eliminate a wide range of possible combinations of internal rotation rates, which would have implications for constraining dynamo models. 

When applied to the measurement of differential rotation of individual stars, the intrinsic limitations of asteroseismology will likely not be overcome in the immediate future. An alternative approach is to perform ensemble fits to a selection of stars of the same type (e.g., Sun-like), thereby placing average limits on the differential rotation rates of that sample \citep{Schunker2016a}. Based on the limits of the radial differential rotation found in this work ($\sim0.4\muHz$), a reasonable sample size to average over would be on the order of a few hundred stars, all with similar properties such as age, surface gravity, effective temperature, and rotation rate. This approach would require a large sample of stars of each type, observed over a long period of time and at high cadence. The \kepler sample only contains a few hundred main-sequence stars observed with such criteria, spanning a range of spectral types. However, with the future launch of the ESA PLATO mission \citep{Rauer2014} this sample will grow to thousands of main-sequence stars. 

\begin{acknowledgements}
M.B.N., H.S., L.G., and W.H.B. acknowledge research funding by Deutsche Forschungsgemeinschaft (DFG) under grant SFB 963/1 ``Astrophysical flow instabilities and turbulence'' (Project A18). M.B.N. received funding from the Deutscher Akademischer Austauschdienst (DAAD) through the Go8 Australia-Germany Joint Research Co-operation Scheme. L.G. acknowledges support from the Center for Space Science at the NYU Abu Dhabi Institute under grant G1502. This paper includes data collected by the \kepler mission. Funding for the Kepler mission is provided by the NASA Science Mission directorate. Data presented in this paper were obtained from the Mikulski Archive for Space Telescopes (MAST). STScI is operated by the Association of Universities for Research in Astronomy, Inc., under NASA contract NAS5-26555. Support for MAST for non-HST data is provided by the NASA Office of Space Science via grant NNX09AF08G and by other grants and contracts.
\end{acknowledgements}

\bibliographystyle{aa}
\bibliography{main.bib}
\end{document}